\documentclass{article}

\usepackage{arxiv}
\usepackage{graphicx}
\usepackage[utf8]{inputenc} % allow utf-8 input
\usepackage[T1]{fontenc}    % use 8-bit T1 fonts
\usepackage{hyperref}       % hyperlinks
\usepackage{url}            % simple URL typesetting
\usepackage{booktabs}       % professional-quality tables
\usepackage{amsfonts}       % blackboard math symbols
\usepackage{nicefrac}       % compact symbols for 1/2, etc.
\usepackage{microtype}      % microtypography
\usepackage{siunitx}

\newcommand{\TP}{\text{TP}}
\newcommand{\TN}{\text{TN}}
\newcommand{\FP}{\text{FP}}
\newcommand{\FN}{\text{FN}}

\title{Staging Epileptogenesis with Deep Neural Networks}

\author{
	Diyuan~Lu\\
	Frankfurt Institute for \\
	Advanced Studies (FIAS)\\
	Frankfurt am Main, Germany\\
	\texttt{elu@fias.uni-frankfurt.de}
	\And
	Sebastian Bauer \\
	Epilepsy Center Frankfurt \\
	Rhine-Main Neurocenter\\
	Frankfurt am Main, Germany\\
	\texttt{Sebastian.Bauer@kgu.de} \\
	\And
	Valentin Neubert \\
	Oscar-Langendorff-Institute \\
	for Physiology\\
	Rostock, Germany\\
	\texttt{valentin.neubert@uni-rostock.de} \\
	\And
	Lara Sophie Costard \\
	Royal College of \\
	Surgeons Ireland\\
	Dublin, Ireland\\
	\texttt{laracostard@rcsi.com} \\
	\And
	Felix Rosenow \\
	Epilepsy Center Frankfurt \\
	Rhine-Main Neurocenter\\
	Frankfurt am Main, Germany\\
	\texttt{rosenow@med.uni-frankfurt.de} \\
	\And
	Jochen Triesch\\
	Frankfurt Institute for \\
	Advanced Studies (FIAS)\\
	Frankfurt am Main, Germany\\
	\texttt{triesch@fias.uni-frankfurt.de}
}

\begin{document}
	\maketitle
	
	\begin{abstract}
		Epilepsy is a common neurological disorder characterized by recurrent seizures accompanied by excessive synchronous brain activity. The process of structural and functional brain alterations leading to increased seizure susceptibility and eventually spontaneous seizures is called epileptogenesis (EPG) and can span months or even years. Detecting and monitoring the progression of EPG could allow for targeted early interventions that could slow down disease progression or even halt its development. Here, we propose an approach for staging EPG using deep neural networks and identify potential electroencephalography (EEG) biomarkers to distinguish different phases of EPG. Specifically, continuous intracranial EEG recordings were collected from a rodent model where epilepsy is induced by electrical perforant pathway stimulation (PPS). A deep neural network (DNN) is trained to distinguish EEG signals from before stimulation (baseline), shortly after the PPS and long after the PPS but before the first spontaneous seizure (FSS). Experimental results show that our proposed method can classify EEG signals from the three phases with an average area under the curve (AUC) of 0.93, 0.89, and 0.86. To the best of our knowledge, this represents the first successful attempt to stage EPG prior to the FSS using DNNs.
	\end{abstract}

	% keywords can be removed

	\section{Introduction}
	% epilepsy affects a lot of people -- early identifying EPG or even staging EPG could allow for better effective treatment -- we don't fully understand the EPG, the progression -- there are attempts in discovering EPG biomarkers, we previously tried to identify BL and EPG -- complexity of EPG, it is not like a single biomarker could be responsible -- thanks to recent advances in ML and DL, we are able to learn the features directly in an end-to-end fashion. -- Here, we propose a DNN-based approach. We 
	Epilepsy is one of the most common and disruptive neurological disorders affecting about 1\% of the world's population. It is characterized by recurrent unprovoked seizures and is accompanied by various co-morbidities such as migraine, depression, dementia, etc. \cite{keezer2016comorbidities}. Over 30\% of the patients will eventually develop refractory epilepsy, defined as inadequate control of seizures by any medication \cite{kwan2000early}. In acquired epilepsy, an initial precipitating injury (IPI) such as stroke, traumatic brain injury or encephalitis leads to structural and functional remodelling of neuronal networks resulting in the occurrence of spontaneous seizures after a clinically silent latent period \cite{pitkanen2014past}. This remodelling process is termed epileptogenesis (EPG). Traditionally, epilepsy is diagnosed and treated after at least one unprovoked seizure, which indicates that the EPG has already progressed to a relatively advanced stage. This latent period can last months or even years. Treating high-risk patients at the early stage of EPG, or even customizing the treatment based on the severity of EPG could result in more effective disease-altering or even disease-arresting outcomes. 
	
	%Addressing this problem is challenging because we don't know much.  
	Pathomechnisms of EPG are not fully understood and its detection remains a major challenge. Studying early EPG in human patients is extremely difficult, simply because the epilepsy is typically only detected after the FSS. Therefore, work on early EPG is typically restricted to animal models \cite{becker2018animal}. Furthermore, early EPG can comprise a complex cascade of changes to the brain after the initial brain insult and this cascade may strongly depend on the type of brain insult. Changes can include, e.g., inflammatory reactions or blood-brain-barrier damage  \cite{engel2020biomarkers}. Some of these brain changes may be reflected in the EEG in the form of interictal epileptiform discharges (IEDs, including sharp-waves, spikes, spike-and-waves complex.), high-frequency oscillations, slowing or alteration of sleep spindles.  Correspondingly, there have been attempts to identify suitable EEG biomarkers for EPG using a wide range of approaches \cite{bragin2004high,li2018extrahippocampal,milikovsky2017electrocorticographic,andrade2017generalized,bentes2018early,lu2019deep, rizzi2019changes}. However, a reliable staging of EPG based on EEG measurements has not been demonstrated yet to the best of our knowledge.
	
\begin{figure}[tb]
		\centering
		\includegraphics[width=0.8\linewidth]{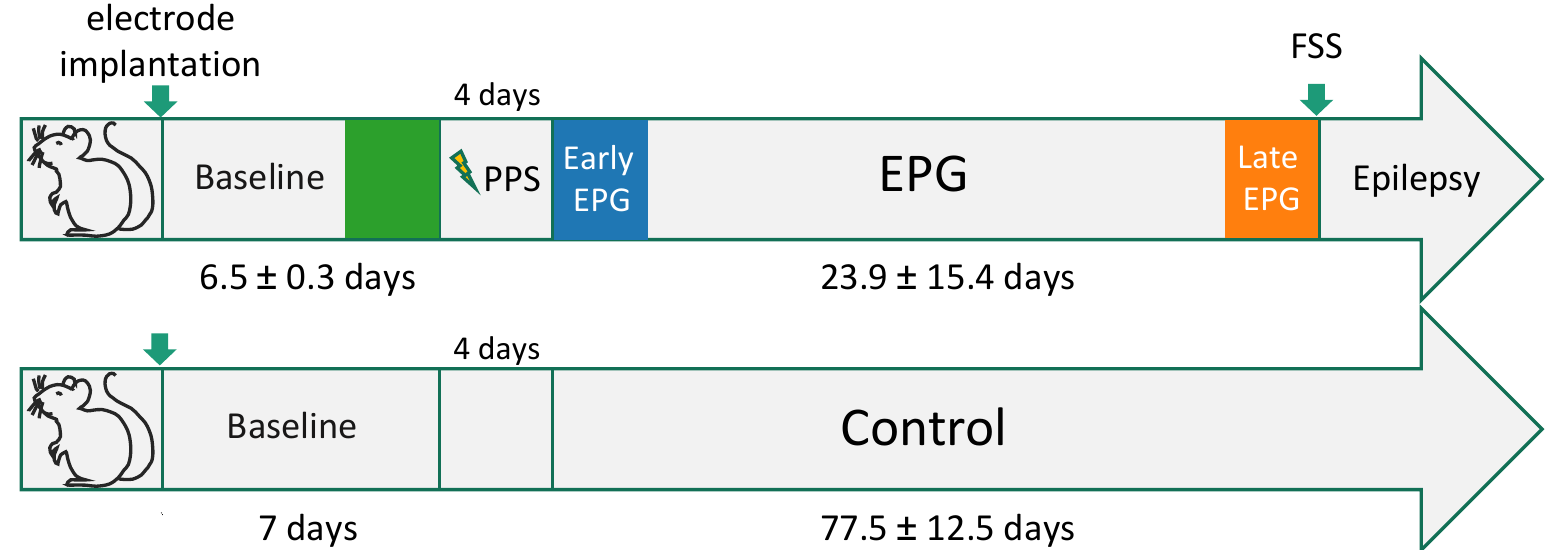}
		\caption{Timeline of the experiment. Shaded boxes indicate the different time periods where training and testing data are extracted. Upper row: PPS group. Lower row: control group (identical but without PPS). FSS: First Spontaneous Seizure. PPS: perforant pathway stimulation.}
		\label{timespan}
	\end{figure}

	Here, we use a rat epilepsy model, where EPG is induced by electrical perforant pathway stimulation (PPS) \cite{costard2019electrical}. In previous work, we have shown that a DNN can be trained to distinguish EEG signals from baseline and EPG, i.e., before and after the PPS, with high specificity and sensitivity. Furthermore, we have demonstrated generalization to unseen rats \cite{lu2019deep}. Here we extend these results and present the first attempt to stage EPG using DNNs. In particular, we ask whether a DNN can also learn to distinguish early and late phases of EPG after the PPS but prior to the FSS, thereby allowing to estimate how ``close'' an individual may be to their FSS. The timeline of the experiment is shown in Fig~\ref{timespan}. There are two groups of rats involved: a PPS group and a control group. The PPS group undergoes PPS and develops epilepsy before the end of the recording. The control group is not stimulated and they do not develop epilepsy before the end of the recording. Data from the control rats are used as a comparison to the PPS group. We demonstrate that our approach based on DNNs can successfully stage the EPG process and distinguish early from late EPG and that it generalizes to previously unseen rats.

	\section{Related Work}
	
	\subsection{Deep Learning for EEG analysis}
	Deep Learning (DL) techniques are commonly used in the analysis of EEG data in medical research. Example applications include the detection of Alzheimer's disease  \cite{bi2019early}, autism \cite{bosl2018eeg}, or Parkinson's disease \cite{faghri2018predicting}. In the context of epilepsy, DL has been applied for abnormal brain activity detection \cite{roy2019chrononet, tjepkema2018deep} as well as seizure detection and prediction \cite{zhou2018epileptic, kiral2018epileptic, thodoroff2016learning, tzimourta2018epileptic, cho2020comparison}.
	Roy \textit{et al.} proposed a hybrid CNN and gated recurrent units (GRU) in classifying normal and abnormal brain activity, which takes time series EEG data as input and outputs the probability of being normal and abnormal, which is one of the first steps to understand the state of the brain activity in order to improve the accuracy of the diagnosis and the quality of patient care \cite{roy2019chrononet}.
	Tjepkema \textit{et al.} explored different combinations of CNNs and recurrent neural networks (RNNs) as classifier to identify IEDs from scalp EEG  \cite{tjepkema2018deep}.
	Zhou \textit{et al.} proposed a CNN-based approach to classify EEG time series data from different states, i.e., ictal, preictal, and interictal for the purpose of seizure detection  \cite{zhou2018epileptic}. They also compared the performance with time series and frequency-domain as input and found that frequency-domain input exhibits better potential for this task. Kiral-Kornek \textit{et al.} proposed a DL-based approach for patient-specific seizure prediction by classifying intracranial EEG data in pre-ictal and interictal phases \cite{kiral2018epileptic}.
	Thodoroff \textit{et al.} proposed a neural network combining convolutional layers (conv-layers) with recurrent layers to detect seizure onset. Their network takes the image-based representation of EEG signals as input capturing spatial, spectral, and temporal features of patient-specific seizures \cite{thodoroff2016learning}. 
	Cho \textit{et al.} compared the performance of different input modalities of EEG data with different DNN-based network architectures for seizure detection \cite{cho2020comparison}. They concluded that the CNN with time-series EEG data, and the RNN with periodogram data resulted in the best performance. While these works have demonstrated the utility of DL for EEG analysis in the context of epilepsy, they have not addressed the challenging detection and staging of EPG prior to the FSS that we demonstrate here for the first time.

	\subsection{Interpretable DNNs}
	The interpretation of the reasoning of a neural network is crucial in medical applications, as it allows verification by human users and provides insights rather than just succumbing to a \textit{black box}. Many studies have been done to address the interpretability of DNNs \cite{simonyan2017deep,sturm2016interpretable,Yosinski2015a,SebastianOn,Kindermans2018, zhou2016learning}. Yosinski \textit{et al.} developed a software tool for visualizing live feature extraction in the neural network by viewing the activation maps of different channels in different layers as well as by regularized optimization to generalize inputs that maximize the channel activation \cite{Yosinski2015a}. Simonyan \textit{et al.} proposed to generate an input image that maximizes the output softmax probability of a given class. Meanwhile, a saliency map can be computed, which is the ranking of each pixel based on their contribution to the given class of a given sample \cite{simonyan2017deep}. 
	Bach \textit{et al.} proposed the Layer-wise Relevance Propagation (LRP), which understands the learning of the network by decomposing the output in terms of the input dimensions in a fashion that relates to Taylor decomposition \cite{SebastianOn}. Sturm \textit{et al.} applied the LRP technique to visualize the frequency contribution to the classification result with EEG data \cite{sturm2016interpretable}. Zhou \textit{et al.} proposed the concept of class activation map (CAM), which can identify important regions in the inputs by propagating back the weights of the dense softmax layer to the inputs \cite{zhou2016learning}. CAM is easy to deploy and provides more focused and localized discrimination. In this work, we also leverage CAM with 1-$d$ EEG data to better visualize the network properties and the learned features.

	% biomarker of EPG from EEG
	\subsection{EEG-based Biomarkers of Epileptogenesis}
	Over the last decades several studies have attempted to find EPG biomarkers in EEG signals. Li \textit{et al.} and Bragin \textit{et al.} focused on high-frequency oscillations (HFOs) in a rat epilepsy model with kainic acid (KA) injection \cite{bragin2004high, li2018extrahippocampal}. They found that the sooner HFOs appear after the injection, the higher the rate of spontaneous seizures in the chronic phase, and the shorter the latent period is, the more spontaneous seizures will occur. Milikovsky \textit{et al.} focused on theta band activity and showed that a decreased theta power can be a robust feature in identifying EPG in five animal epilepsy models \cite{milikovsky2017electrocorticographic}.
	Andrade \textit{et al.} investigated the role of sleep-wake disturbance in EPG and found that there is a decrease of the dominant frequency and the duration of sleep spindles in a traumatic brain injury epilepsy model with generalized seizures \cite{andrade2017generalized}. 
	Bentes \textit{et al.} found that in stroke patients, the asymmetry in the background activity with the occurrence of IEDs are independent indicators of post-stroke epilepsy in the first year after stroke \cite{bentes2018early}.
	Sheybani \textit{et al.} found that in a mouse model of epilepsy with kainate injection, the spatial propagation of a subgroup of spikes across the brain can be a reliable indicator of EPG as well as epilepsy in the chronic phase \cite{sheybani2018electrophysiological}.
	Lu \textit{et al.} trained a DNN with the Fourier transformation of the time-series EEG data from a rat epilepsy model and showed that a decrease of power in theta band and an increase of power in frequencies over 100 Hz can be reliable indicators of EPG \cite{lu2019deep}.
	Rizzi \textit{et al.} investigated the nonlinear dynamics of EEG signals and found a significant decrease of the so-called embedding dimension in early EPG that correlates with the severity of the ongoing EPG \cite{rizzi2019changes}.
	Here, we use an unbiased deep learning approach to study the EPG process to subdivide it into different stages and identify potential biomarkers to distinguish early and late phases of EPG.

	% Since the control rats never had PPS we used the following processing steps to make the time window comparable with the PPS group. We denote the first seven days after the implantation as BL, then we skip four days, where the PPS group underwent PPS and denote the remaining period as \textit{pseudo EPG} as opposed to \textit{EPG} in the PPS group.

	\section{Methods}
	% rewrote
	\subsection{Animal Model}
	We use a mesial temporal lobe epilepsy with hippocampal sclerosis (mTLE-HS) rodent model, where epilepsy is electrically induced through PPS. Details have been described in \cite{costard2019electrical}. Continuous single-channel EEG recordings from a depth electrode implanted in the dentate gyrus are collected from each rat from the beginning of the implantation until the FSS, which indicates the manifestation of epilepsy. The 24/7 recordings enable us to continually monitor the entire EPG prior to the FSS. There are two groups of rats involved in this study, 1) seven rats had PPS and developed epilepsy before the end of recording, which we denote as PPS rats, 2) three rats did not get PPS stimulation and did not develop epilepsy by the end of recording, which we denote as control rats. In the PPS group, the average EPG phase is 4 weeks (range 1 -- 7 weeks). The EPG phase is terminated by the FSS. The timelines for both group are shown in Fig.~\ref{timespan}. Training data are taken from the three highlighted periods from PPS rats for the three-class classification task. We define the three classes to be the Baseline class (\textit{BL}) -- green, the \textit{early EPG} class -- blue, and the \textit{late EPG} class -- orange. The data from the control rats are used only for testing the model trained on the PPS group. The total available number of recordings from each rat is summarized in Table~\ref{tab:PPS} and Table~\ref{tab:control}.
	
	\begin{table}
		\centering
		\caption{Summary of the data collections from \textbf{PPS} rats in hours (hrs).}
		\label{tab:PPS}
		\begin{tabular}{cccccccc}
			\toprule
			rat ID & PPS 1 & PPS 2 & PPS 3 & PPS 4 & PPS 5 & PPS 6 & PPS 7 \\
			\midrule
			BL (hrs) & 162 & 160& 149& 82& 163 & 164 & 157\\
			EPG (hrs) & 700  & 508 & 400 & 140 & 1568 & 173 & 648\\
			\bottomrule
		\end{tabular}
	\end{table}
	
	\begin{table}
		\centering
		\caption{Summary of the data collections from \textbf{control} rats in hours (hrs).}
		\label{tab:control}
		\begin{tabular}{cccc}
			\toprule
			rat ID & Ctr 1 & Ctr 2 & Ctr 3 \\
			\midrule
			in total (hrs) & 1536  & 2140 & 2248\\
			\bottomrule
		\end{tabular}
	\end{table}
	
	\subsection{EEG Data Preprocessing}
	The data acquisition was achieved through wireless EEG transmitters with a sampling rate of 512 Hz and a band-pass filter between 0.5 - 160 Hz as well as a notch filter at 50 Hz. Occasionally, EEG artifacts can appear as extreme amplitude values and signal loss due to electronic interference and weak transmission. To combat this problem,  we first applied a MATLAB function, i.e., \texttt{filloutliers} \footnote{\url{https://www.mathworks.com/help/matlab/ref/filloutliers.html}} with the parameters \texttt{method = 'pchip'; movmethod = 'movmedian'; window = 50} to filter out unrealistic extreme values. Then, the continuous recordings are divided into five-second long non-overlapping segments. To manage data loss, we discarded any five-second segments with more than 20~\% data loss. As a result, we discarded around 5\% of the total recordings. The remaining segments were eligible for the DNN training. 
	% {\bf JT: What fraction is discarded? What fraction are really clean, i.e., completely without data loss?}

	\subsection{DNN Architecture}
\begin{figure*}[tb]
		\centering
		\includegraphics[width=0.85\linewidth]{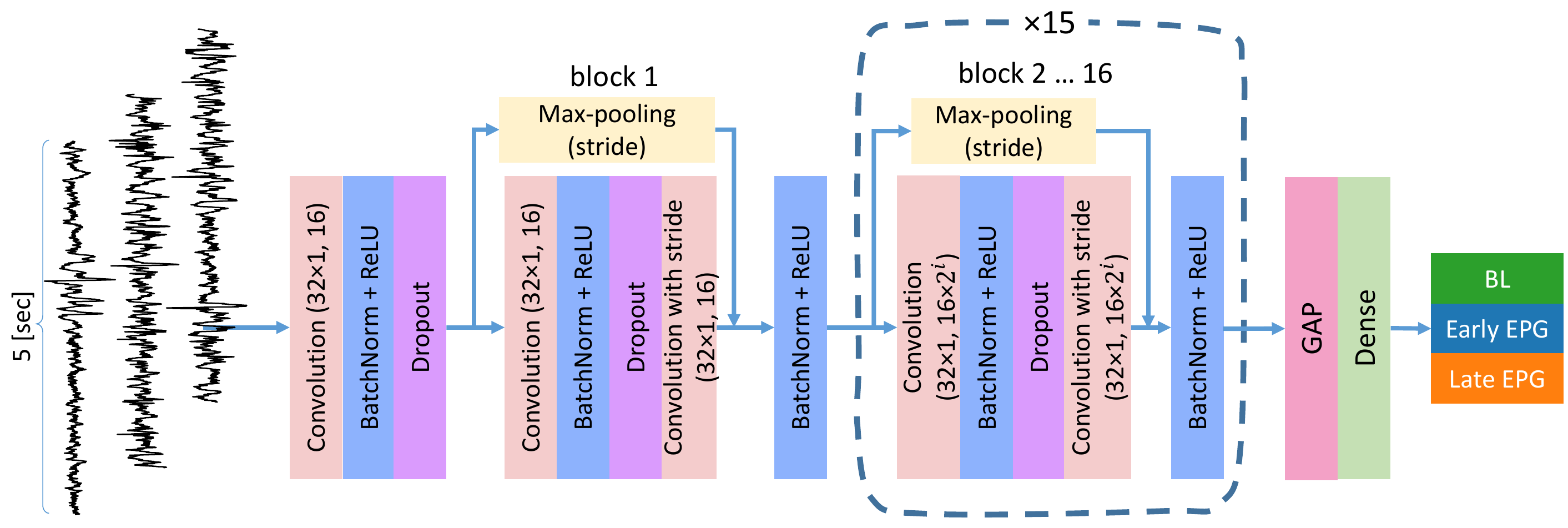}
		\caption{The DNN structure used in this study. The network takes a mini-batch of five-second segments as input and outputs the probability over the three classes. GAP: global average pooling. BL: Baseline}
		\label{fig:Whole_structure}
	\end{figure*}

	We use a deep residual neural network with 16 blocks with residual connections (res-block) as shown in Fig.~\ref{fig:Whole_structure}, inspired by \cite{hannun2019cardiologist}. The model takes five-second long EEG segments as input and outputs the probability over three classes, i.e., BL, early EPG, and late EPG. 
	We keep the design of each res-block as in \cite{hannun2019cardiologist}, where each res-block consists of two conv-layers, batch-normalization, dropout, and ReLU non-linear activation. The number of channels in the first conv-layer and the first block is 16, and it increases by a factor of 2 in every four blocks. There are two branches in each block: one goes through convolution, batch-normalization, ReLU activation and dropout; the other, called skip connection, simply goes through max-pooling. They are combined in an additive manner at the end of the block before passing through the batch-normalization and ReLU activation. To reduce the dimensionality of the feature maps, we use a stride of two in the second conv-layer and the max-pooling layer in every other block starting from the second block. The output of the last conv-layer is fed to the global average pooling (GAP) operation, which is followed by a dense layer with three output units with softmax non-linear activation. The dropout rate is 0.2 everywhere in the graph.
	
	\subsection{Class Activation Map}
	Proposed by Zhou \textit{et al.}, the class activation map is a method to visualize the ``importance'' of different regions of the input for the classification decision. It takes advantage of the global average pooling (GAP) after the last conv-layer, and assigns different weights to each squashed feature map. To be specific, the $k$-th feature map from the last conv-layer, denoted as $f_k$, which has shape [$height$, $width$]. The GAP layer takes the mean activation of each $f_k$, and the resulting $k$-th feature map $F_k$ is $\frac{1}{N}\sum_{i,j}f_k(i,j)$, where $N$ is the total number of elements of $f_k$. It reduces the dimension by the factor of $height \times width$. Then, for a given class $c$, the input to the softmax layer, $S_c$, is a weighted linear combination of all the feature maps, which is computed by
	\begin{equation}\label{eq:y_c}
		S_c = \sum_{k}w^c_k\frac{1}{N}\sum_{i,j}f_k(i,j) \propto \sum_{i,j}\sum_{k}w^c_kf_k(i,j),
	\end{equation}
	where $w^c_k$ denotes the \textit{importance} of $f_k$ for class $c$. Finally, the softmax probability for class $c$ can be computed as $\frac{\exp{S_c}}{\sum_c\exp{S_c}}$. 
	Then, when the training is finished, the class activation map for class $c$ at position ($i, j$), $\text{CAM}_c (i,j)$, is given by \begin{equation}
		\text{CAM}_c (i, j) = \sum_kw^c_kf_k(i, j).
	\end{equation}
	Hence, $S_c = \sum_{i,j}w^c_k\text{CAM}_c (i, j)$, and the weights $w_c$ are fixed after the training. Then, $\text{CAM}_c(i,j)$ indicates the importance of the activation at the position $(i, j)$ contributing to the class $c$.
	
	\subsection{DNN Training and Evaluation}
	We apply a seven-fold leave-one-out cross-validation (LOO-CV) scheme, where the network is trained with the data from six out of seven rats in the PPS group. The data from the last rat are held out as the test set. This procedure is repeated seven times, and each time we hold out a different rat for testing. This is highly relevant to test the generalization ability of the classifier to unseen data from unseen subjects. We randomly select 25 hours from a three-day window from each phase for training and validation, shown as the shaded periods in Fig.~\ref{timespan}. The choice of 25 hours is a reasonable trade-off between computational cost and performance from empirical experience. Our DNN model is implemented in Tensorflow and trained with an NVIDIA GeForce RTX 2080 Ti GPU. 
	Among all the selected data for training and validation, we adopt a train-validation-split of 8:2. After the network is trained, we test it with all the data from those three-day periods (shown in Fig.~\ref{timespan}) of the previously withheld rat. We report results as the average across all seven LOO test trials.
	% \begin{figure}[bt]
	%   \centering
	%   \includegraphics[width=0.9\linewidth]{LOOcv.pdf}
	%   \caption{Leave-one-out cross validation scheme in our study. In each trial, we use the data from six rats with a 8:2 train-validation split and hold out the data from the last rat for LOO testing.}
	%   \Description{LOO-scheme}
	%   \label{fig:LOOcv}
	% \end{figure}

	To evaluate the performance, we compute the receiver operating characteristic (ROC) curve in the multi-class scenario, where the ROC curve is computed for each class in a one-vs-all manner. The area under the ROC curve is a scalar value indicating the goodness of the trained classifier. Several other performance metrics including precision, recall, and F1-score are also computed. These metrics are given by:
	\begin{eqnarray*}
		\text{precision} & = & \frac{\TP}{\TP + \FP} \\ 
		\text{recall}    & = &  \frac{\TP}{\TP + \FN} \\
		\text{F1-score}  & = & 2 \cdot \frac{ \text{precision} \cdot \text{recall}}{\text{precision} + \text{recall}}  \\
		\text{accuracy}  & = & \frac{\TP + \TN}{\TP + \TN + \FP + \FN},
	\end{eqnarray*}
	where \TP, \TN, \FP, and \FN~ are true positive, true negative, false positive, and false negative numbers, respectively.
	We also compare our results with several baseline network structures: a feed-forward neural network (FNN), a deep convolutional neural network (DCNN) \cite{sors2018convolutional}, EEGNet \cite{lawhern2018eegnet}, and one variant of our proposed model with only four blocks, which we denote as Proposed-4block. 
	
	The FNN used in this work is a straight forward multi-layer perceptron with three dense layers equipped with 1024, 256, and 128 units per layer. Each dense layer is regularized with $L_2$ penalty with a factor of 0.01 and followed by a batch-normalization layer and a dropout (rate=0.5) layer. The nonlinear activation is ReLU in this model.
	
	Sors \textit{et al.} proposed the DCNN for sleep staging with single-channel EEG. Compared to the original architecture, we made several changes to adapt to the training data format we have in our experiment. First, due to our input being shorter (five-second segments under 512 Hz sampling rate, which yields 2560 data points per sample) than theirs (\num{15000} data points), we reduced the number of conv-layer from twelve to nine: five (instead of six) conv-layers with 128 output channels and four (instead of six) conv-layers with 256 output channels. Each conv-layer has stride 2 to sub-sample the feature map. The architecture is conv ($7\times 1, 128$, stride 2) -- conv ($7\times 1, 128$, stride 2) -- conv ($7\times 1, 128$, stride 2) -- conv ($7\times 1, 128$, stride 2) -- conv ($7\times 1, 128$, stride 2) -- conv ($5\times 1, 256$, stride 2)  -- conv ($5\times 1, 256$, stride 2)  -- conv ($5\times 1, 256$, stride 2)  -- conv ($3\times 1, 256$, stride 2) -- flatten -- fully-connected (units=100) -- fully-connected (units=3). We kept other training parameters identical to the original paper.
	
	\begin{table*}[t]
		\centering
		\caption{Performance measures across all leave-one-out test trials with one hour of prediction aggregation. Evaluation metrics are reported in class-wise average and overall average for each model. Numbers are shown in  $mean \pm std $.}
		\centering
		\begin{tabular}{ccccccc}
			\toprule
			\textbf{Model} & \textbf{Class} & \textbf{Precision}  & \textbf{Recall} & \textbf{F1-score} & \textbf{Accuracy} & \textbf{\# trainables}\\
			\midrule				
			FNN  & 0 & $0.51 \pm 0.08$  & $0.67 \pm 0.13$ & $0.57 \pm 0.06$ &  $0.44 \pm 0.06$ & \num{2920963}\\
			& 1 & $0.49 \pm 0.07$  & $0.65 \pm 0.07$ & $0.55 \pm 0.02$ &  $0.43 \pm 0.04$   &\\
			& 2 & $0.43 \pm 0.06$  & $0.63 \pm 0.16$ & $0.49 \pm 0.04$ &  $0.39 \pm 0.03$   &\\
			& average & $0.47 \pm 0.03$  & $0.65 \pm 0.04$ & $0.53 \pm 0.03$ &  $0.42 \pm 0.03$  & \\
			\midrule
			DCNN     & 0 & $0.47 \pm 0.13$  &  $0.54 \pm 0.25$ &  $0.46\pm 0.18$ &  $0.46\pm 0.13$  &  \num{1607187}\\
			\cite{sors2018convolutional} & 1 & $0.43 \pm 0.32$  & $0.40 \pm 0.30$  & $0.41 \pm 0.31$  & $0.41 \pm 0.08$ &\\
			& 2 &  $0.35 \pm 0.23$ & $0.35 \pm 0.28$  &$0.33 \pm 0.23$  &  $0.40 \pm 0.07$  &\\
			& average &  $0.42 \pm 0.11$  &  $0.43 \pm 0.22$ & $0.40 \pm 0.18$   & $0.42\pm 0.07$ & \\
			\midrule
			Proposed-4block & 0  & $0.70 \pm 0.14$  & $0.88 \pm 0.04$ & $0.78 \pm 0.10$ &  $0.66 \pm 0.14$ & \num{82912}\\
			& 1 & $0.43 \pm 0.06$  & $0.68 \pm 0.18$ & $0.53 \pm 0.10$ &  $0.41 \pm 0.05$   &\\
			& 2 & $0.51 \pm 0.05$ &  $\textbf{0.82} \pm \textbf{0.13}$& $0.62 \pm 0.01$ & $0.47 \pm 0.02$ &\\
			& average &  $0.55 \pm 0.04$ & $0.79 \pm 0.09$& $0.64 \pm 0.05$ & $0.51 \pm 0.05$ & \\
			\midrule         
			Proposed model & 0 & $\textbf{0.85} \pm \textbf{0.17}$  & $\textbf{0.96} \pm \textbf{0.02}$  & $\textbf{0.90} \pm \textbf{0.10}$  & $\textbf{0.84}\pm \textbf{0.17}$ & \num{4200048}\\
			& 1 & $\textbf{0.69} \pm \textbf{0.12}$  &  $\textbf{0.81} \pm \textbf{0.17}$ &  $\textbf{0.74} \pm \textbf{0.14}$ &  $\textbf{0.64} \pm \textbf{0.15}$  &\\
			& 2 &  $\textbf{0.71} \pm \textbf{0.33}$ & $0.74\pm 0.32$  &$\textbf{0.72} \pm \textbf{0.31}$  &  $\textbf{0.71} \pm \textbf{0.22}$  &\\
			& average &  $\textbf{0.75} \pm \textbf{0.15}$  &  $\textbf{0.84} \pm \textbf{0.12}$ & $\textbf{0.78} \pm \textbf{0.14}$   & $\textbf{0.73} \pm \textbf{0.14}$ & \\
			\bottomrule
		\end{tabular}
		\label{tab:performance}
	\end{table*}

	Lawhern \textit{et al.} proposed the original EEGNet for EEG classification in multiple brain-computer interfaces. The EEG snippets used in their evaluation are multi-channel event related potential (ERPs) recorded from surface EEG setups, band-pass filtered between 1-40 Hz, downsampled to 128 Hz, and focused on 1 to 2 seconds around the event onset. The original EEGNet demonstrates good generalization to EEG classification among different experiment diagrams even though the total number of parameters is two orders of magnitude smaller than the baseline methods evaluated in their work. To adapt EEGNet to our task, we made several changes to the architecture while keeping layers such as batch-normalization, dropout, exponential linear unit (ELU) activation function, and average pooling unchanged: 1) We expanded the width of the convolutional filter from 64 to 256, which is half of our sampling rate as suggested in the original paper. 2) We used three instead of two layers of convolution while omitting the depth-wise convolution, since our data is single-channel. Unfortunately, the classification accuracy of this modified EEGNet (henceforth denoted EEGNet1) does not exceed chance-level. One contributing factor might be the low number of trainable parameters. In total, EEGNet1 only has \num{223323} learnable parameters, which is considerably fewer than our proposed model. To make the total number of trainable comparable to ours, we increased the number of conv-layers and the number of filters in each layer. This is essentially equivalent to a relatively shallow CNN (7 conv-layers compared to 33 layers in our proposed model) with very wide convolutional filters, which we denote as EEGNet2. The resulting structure of EEGNet2 is conv ($256\times 1, 16$) -- batch-normalization -- conv ($256\times 1, 16$) -- batch-normalization + ELU + average-pooling + dropout -- conv ($256\times 1, 32$) -- batch-normalization + ELU + average-pooling + dropout -- conv ($256\times 1, 32$) -- batch-normalization + ELU + average-pooling + dropout -- conv ($256\times 1, 64$) -- batch-normalization + ELU + average-pooling + dropout -- conv ($256\times 1, 64$) -- batch-normalization + ELU + average-pooling + dropout -- flatten -- fully-connected (units=3). As a result, the EEGNet2 has a total number of \num{4195107} parameters, which is comparable to that of our proposed model (\num{4200048}). However, the results show that with the same amount of training data and training time, both versions of EEGNets, i.e., EEGNet1 and EEGNet2 perform at chance-level. Thus, their performance measures were omitted in the performance report.
	
	\section{Experiments and Results}
	\label{results}
	
	Table~\ref{tab:performance} shows the performance summary of our proposed model in comparison to the baseline methods. The reported performance metrics are averaged for each class as well as a macro-average of all classes across all LOO test trials. Our proposed method obtains the best performance in almost all evaluated metrics compared to the baseline methods. Notably our proposed-4block model still obtains better performance than FNN and DCNN, even though the number of trainable parameters is more than 20 times smaller. Compared to the full-size proposed model, the Proposed-4block model suffers from a slight performance reduction. From the class-wise performance, we can see that, in general, the BL class is easier for the networks to classify as shown by the highest average performance among the three classes in all models.
	
	\begin{figure*}[th]
		\centering
		\includegraphics[width=0.99\linewidth]{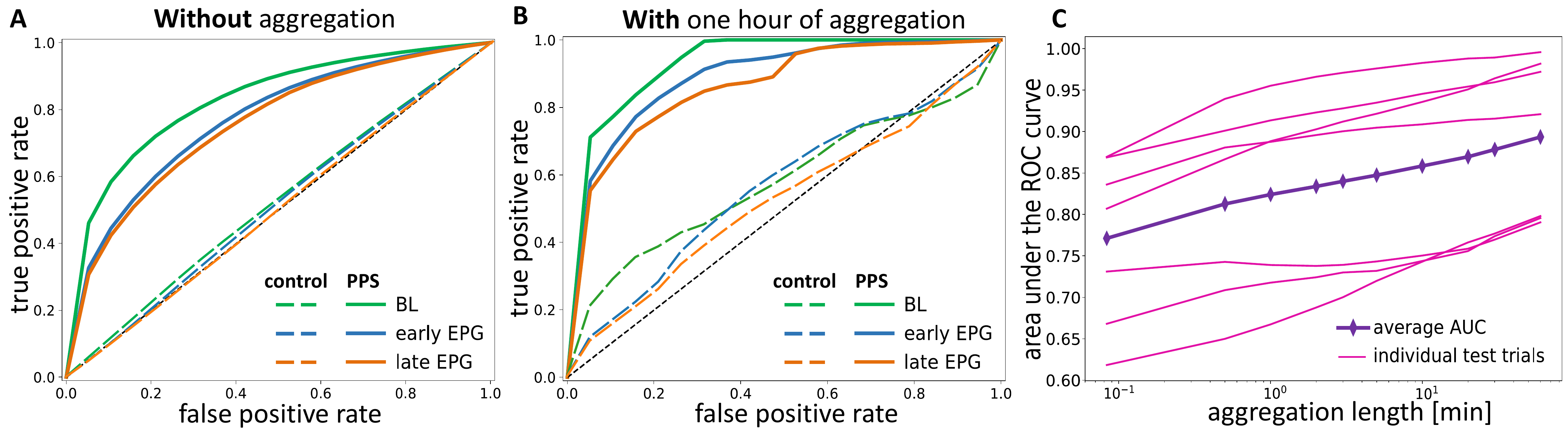}
		\caption{Network performance across all test trials within the PPS and the control group. A. Average ROC curves of multiple classes \textbf{without} aggregation within the PPS group and the control group. The AUC for the three classes of PPS rats are 0.83, 0.77 and 0.75 (solid lines) and those of the control rats are 0.52, 0.51, and 0.50 (dashed lines). B. Average ROC curves of multiple classes \textbf{with} aggregation over one continuous hour within the PPS and the control group. The AUC of the three classes for the PPS rats (solid lines) are 0.93, 0.89, and 0.86, and those of the control rats are 0.58, 0.56, and 0.53 (dashed lines). C. The AUC as a function of the aggregation length in all individual PPS LOO test trials (magenta lines) and the average AUC of all classes across all trials (purple with diamonds).  ROC: receiver operating characteristic. AUC: area under the curve.}
		\label{fig:AUCs}
	\end{figure*}
	
	\begin{figure*}[th]
		\centering
		\includegraphics[width=0.99\linewidth]{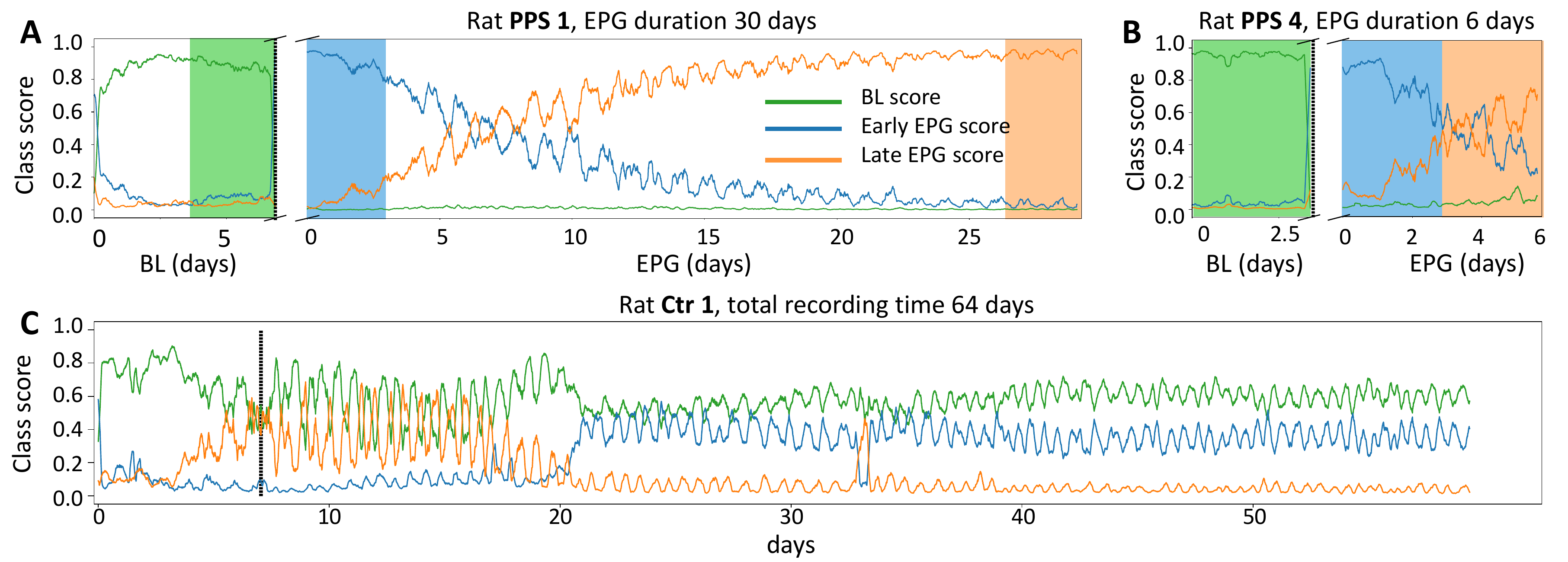}
		\caption{Class scores from two \textbf{PPS} rats (A,B) and one \textbf{control} rat (C) during the entire recording. The vertical black dashed lines indicate the time when the PPS rats started receiving PPS, while control rats did not.}
		\label{fig:class-score}
	\end{figure*}
	
	\subsection{Prediction Aggregation and ROC Analysis}
	To gather statistics of the estimated class membership over a longer time period, we apply a prediction aggregation technique as proposed in our previous study \cite{lu2019deep}. Essentially, we apply a linear average aggregation of the resulting softmax probability across multiple consecutive five second data segments such that the probabilities of each class are accumulated across a longer period of time. Figure~\ref{fig:AUCs} shows the averaged AUCs of the three classes across all LOO test trials \textbf{with} and \textbf{without} the prediction aggregation (Fig.~\ref{fig:AUCs}A and Fig.~\ref{fig:AUCs}B) as well as the effect of the pooling length used in the prediction aggregation (Fig.~\ref{fig:AUCs}C). In general, the network can distinguish BL segments better than the other two classes as shown by the highest average AUC under the ROC curve among the three classes, with or without the prediction aggregation. Prediction results for the control group are only marginally better than chance, suggesting that the network really detects changes in brain activity patterns due to the PPS, rather than any changes triggered by the initial electrode implantation that are independent of the PPS. Prediction aggregation over one hour increases the average AUC of the baseline, early, and late EPG classes by 0.1, 0.12, and 0.11, respectively.

	\begin{figure*}[t]
		\centering
		\includegraphics[width=0.99\linewidth]{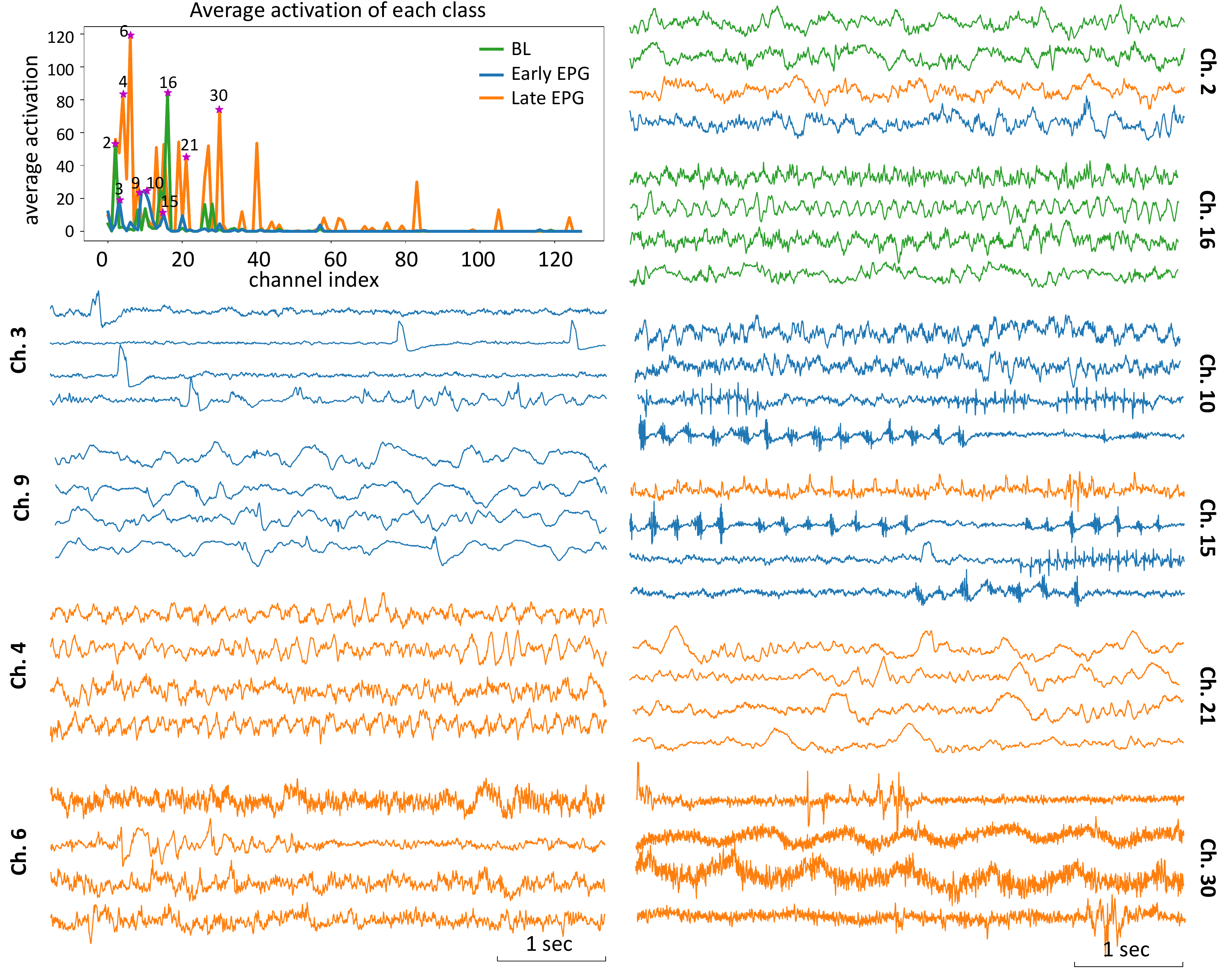}
		\caption{Normalized average activation of the last conv-layer by class (top left). Examples of five second EEG samples that maximize the activation of certain channels in the last conv-layer. Color indicates a sample's class label. Scale bar represents 1 second. 
			% 	{\bf JT: there's no axis label on the x-axis of the samples indicating that it represents time measured in seconds. As an alternative, I suggest to also get rid of the labels 1--5 and instead put in a 1 second scale bar in one of the traces.}
		}
		\label{fig:b4softmax}
	\end{figure*}
	
	\begin{figure*}[th]
		\centering
		\includegraphics[width=1.0\linewidth]{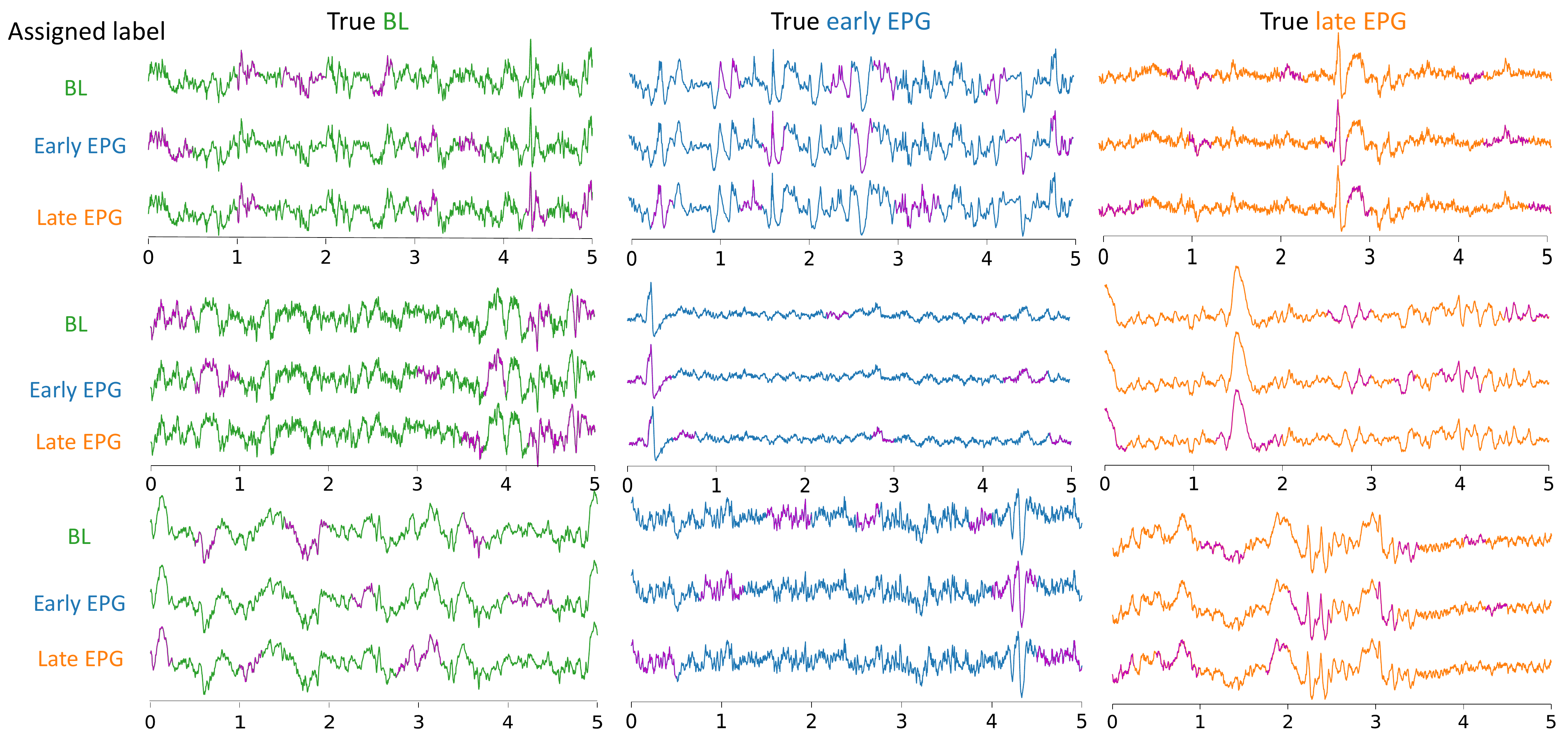}
		\caption{Identifying informative regions in the five-second long input via CAM. The main color of a trace corresponds to its true label. The areas highlighted in magenta most strongly support the assigned classification (>80-th percentile). 
			% 	{\bf JT: You could write the labels at the top in the corresponding color of that class. Instead of the uniformative labels 0,1,2 you could use BL, early, late or sth like that. Also: I'm still not convinced that simply using the 80th percentile is the best choice. Have you looked at a histogram of the actual values? If the distribution is bimodal (or multimodal), then one could use the corresponding split.}
		}
		\label{fig:cams}
	\end{figure*}
	
	To study the benefits of aggregation in more detail, we compute the AUCs for various aggregation lengths in each LOO test trial, i.e., 5 seconds, 30 seconds, one, two, five, ten, 20, 30, and 60 minutes. The average AUC as a function of the aggregation lengths is depicted in Fig~\ref{fig:AUCs}C. It reflects the inter-rat variability in the three-class classification with our proposed network, i.e., the AUC starts at different levels of confidence without prediction aggregation (the first data points from all rats). The figure shows that with an increasing pooling length, the average AUC increases in all LOO test trials. To be specific, with one hour of aggregation, the average AUC improved by 0.12 (a maximum of 0.18 and a minimum of 0.06).  Hence, aggregating the softmax output from the network across multiple consecutive segments captures trends across a longer period, which is essential for distinguishing different classes in our task. Aggregation over even longer time periods (>1 hour) might be able to further improve performance.

	\subsection{Disease Progression}
	EPG is a gradual process, but above we treated EPG detection and staging as a discrete classification problem by defining (somewhat arbitrarily) the first three days after the stimulation as the early EPG phase, and the last three days before the FSS as the late EPG phase. The data from the period in between these two phases has not been considered so far. In the following, we analyze samples from this intermediate period and study how the network, which has been trained to distinguish Baseline, early and late EPG phases, will classify them. Specifically, we consider the estimated probability for each class, denoted as the \textit{class score}, throughout the whole recording period from a randomly picked pre-trained model from the LOO cross-validation scheme, which we call "Pretrained-1" model. Here, we are interested in the general trend rather than the classification accuracy, so the training data were also included. 
	% {\bf JT: I don't understand what you mean with randomly picked model. Why not use a model trained on all rats?} 
	The progression of class scores from two example \textbf{PPS} rats and one \textbf{control} rat are shown in Fig.~\ref{fig:class-score}. One of the PPS rats (PPS~1) has a relatively long EPG duration (30 days) and the other (PPS~4) has a short EPG duration (6 days). The control rat (Ctr~1) has 64 days of recordings in total. 
	
	Several findings are evident in the data for the PPS rats in Fig.~\ref{fig:class-score}A,B. First, the Baseline score is high during the entire baseline period and drops to small values during the EPG phase. Second, with the beginning of the EPG phase, the early EPG score increases and then gradually decreases towards the late EPG phase. Third, conversely, the late EPG score is low during baseline and the beginning of EPG and then gradually increases towards the late EPG phase. Fourth, in some animals we observe a circadian rhythm in the early and late EPG scores during the transition period between early and late EPG (compare Fig.~\ref{fig:class-score}A). These findings are in sharp contrast to those for the control rats. In their case, the late EPG score remains low throughout the entire recording period, in line with these animals not developing epilepsy during the experiment (compare Fig.~\ref{fig:class-score}C).

	\subsection{Feature Representation}
	The interpretation of EEG signals is always challenging, since they are highly variable --- especially across subjects. Analyzing and understanding the discriminative features learned by a DNN model can give valuable insights as to what distinguishes the classes. This can be particularly helpful in medical applications, where the differences between classes many not be easily spotted --- even by the expert eye. Here, we present the feature representations learned by the network. Using the Pretrained-1 model, we passed the unseen 
	% {\bf (JT: Why unseen? What data do you mean?)} 
	data from all seven rats through the network and computed the average activation of the last conv-layer for each class. Due to the massive amount of data, we randomly sampled 190 hours of recordings from each class to reduce the computational load. We then computed the average activation per class, shown on the top left of Fig.~\ref{fig:b4softmax}. We can see that there is a group of feature channels that are very active. Most importantly some of these feature channels are most active for one class but not the others and some extract features that contribute to more than one class.
	% Second, we identified several channels that were highly active for each class for illustration. Third, we presented the input segments that maximally activate those selected channels, shown in Fig.~\ref{fig:b4softmax}. 
	% such as channel 2 and 16 to the BL class, channel 3, 9, 10, and 15 for the early EPG class, and channel 4, 6, and 30 to the late EPG class. 
	Next, we identified several channels that were highly active for each class and plotted the EEG segments that maximally activate them. Interestingly, we found several feature channels responding to very distinctive features such as spikes in channel 3, spike-and-slow-waves in channel 9, spindles and HFOs in channel 15, theta rhythm in channel 16, delta wave plus low beta in channel 21, etc. From this we can conclude that before the softmax layer, the network has already extracted class-specific features that are clinically meaningful. 
	
	% {\bf JT: I think we can improve the above section by doing several things. First, we should find a way to sort the channels meaningfully. Now they have a random order, it seems. How about ordering them by the class that they vote for the most? We should also find a way to mark the ones where we show the samples. Right now it's difficult to identify, which one is, e.g., number 10, because we only have tick marks for every 20th channel. We could try to add more tick marks or put little symbols or numbers next to some of the peaks. Also: be consistent with
	% ``average'' vs. ``averaged''. Finally, it's not clear to me how you picked the example channels. It seems that the channels that vote strongly for one class but not at all for any of the others should be most interesting. So maybe one could derive some score for each channel that allows to rank them in this sense? But feedback from Sebastian should also be used.}
	
	To further elucidate which parts of the input contribute most to the classification of the different EPG stages, we leverage the CAM visualization while manipulating the assigned labels for the EEG segments. Taking Pretrained-1 model, we freeze the weights and for a given sample, we assign in turn the three different labels. Then, by computing the CAM of the given sample under the assigned label, we trace back which parts of the given five second input segment most support (> 80-th percentile) the assigned classification. The results are shown in Fig.~\ref{fig:cams}. Indeed, the CAMs for the sample vary depending on the given label. There are several interesting features that the network has discovered. First, the BL class is most supported by low-amplitude waves, and many downwards deflections. Second, sharp waves contribute to both EPG classes, but the difference lies in the width. While an early EPG classification is supported by narrow spikes, or spike-like waves, a late EPG classification is supported by somewhat wider sharp waves.

	\section{Conclusion}
	We have proposed a DNN model for single-channel intracranial EEG classification to better understand the progression of epileptogenesis (EPG). Specifically, our aim was to stage the EPG process prior to the first spontaneous seizure (FSS), which could facilitate early intervention {\bf before} an epilepsy becomes manifest. In previous work, we had already shown that a DNN can learn to distinguish EEG data from before and after the epilepsy-inducing stimulation with high discrimination and generalization ability \cite{lu2019deep}. Here, we have sought to answer a) whether we can further distinguish different stages of EPG before the FSS, and b) what EEG features would be representative for each stage. To this end, we have trained a DNN model with five-second EEG segments recorded from three phases in a rodent epilepsy model \cite{costard2019electrical}: three days before the PPS (Baseline, BL), three days shortly after the PPS (early EPG), and three days immediately before the FSS (late EPG). We have evaluated our approach in a LOO scheme to test the generalization ability of the model to data from unseen rats. To pool evidence over larger time windows, we applied a prediction aggregation method as in previous work \cite{lu2019deep}. We also compared the performance of our model to four other models, specifically an FNN model, a DCNN model \cite{sors2018convolutional}, the well-known EEGNet \cite{lawhern2018eegnet}, and a reduced version of our model with 50 times fewer parameters. In an extensive performance evaluation, we showed that our proposed model yielded the best results and could distinguish different EPG stages with high accuracy. Furthermore, we showed that the network learns to extract meaningful EEG features to perform the classification.
	
	%We also tested our model with data from control rats and not surprisingly the network only performs at chance-level, indicating that there exit distinct patterns in the data from PPS group to distinguish the three stages. To visualize the learned features of our model, we provide visualizations from two perspectives. First, we identify several channels in the last conv-layer that are most responsive.
	% Representative EEG segments that maximally activate those channels were collected and presented in Fig.~\ref{fig:b4softmax}. From this figure, we found that the features different channels are looking for are with distinct characteristics. 
	%Second, we leveraged the class activation map by varying labels assigned to the same sample to visualize the contribution of different regions of the input to the final class label more precisely. The visualization of the learned features of the network confirmed that the network has learned meaningful features that contribute to different stages of EPG. In this work, we demonstrated the feasibility of sub-dividing the EPG phase before the FSS with DNNs and presented representative features learned by the network that attribute to different stages.
	
	Various challenges will need to be overcome, in order to translate our findings to human patients. First, the rodent model we have used provides quasi ideal conditions, supplying high quality, 24/7 intracranial recordings directly from the affected brain region. It is unclear whether similar results could be achieved with surface EEG recordings from a diverse set of human patients. The second challenge is that epilepsy is typically diagnosed only {\em after} the FSS. In order to attempt early detection of EPG as we have demonstrated here in human patients, one would have to obtain recordings from patients {\em before} the FSS. This requires monitoring a population of patients with a sufficiently high risk of developing epilepsy, which is challenging. Third, our approach relies on a large data set comprising around-the-clock recordings over several weeks for each individual. Acquiring similar data from a (homogeneous) patient population would be very difficult. It is an open question, how much data would be required to allow accurate classification and good generalization. Fortunately, in our experiments, pooling data over one hour already provided very good results. Such a time span appears manageable in clinical practice. Finally, even if EPG could be detected and staged reliably in human patients at risk of developing epilepsy, it is far from clear which forms of early intervention would be effective in modifying or halting the disease development. In fact, such interventions will likely have to depend on the specific type of epilepsy and be adapted to individual patients. In the future, machine learning may also support physicians in this challenging task.
	
	\section*{Acknowledgement}
	This work is supported by the China Scholarship Council (CSC, No. [2016]3100), the LOEWE Center for Personalized Translational Epilepsy Research (CePTER), and the Johanna Quandt Foundation. The authors would like to thank Markus Ernst for proofreading the paper and providing valuable feedback. 
	
	\bibliographystyle{unsrt}
	\bibliography{ACM-reference}
	

	% \section{Class Scores}
	% Class scores of other rats
	% \begin{figure*}[t]
	% 	\centering
	% 	\includegraphics[width=0.9\linewidth]{class-score-other-rats.pdf}
	% 	\caption{Class scores}
	% 	\label{fig:class-score-other}
	% \end{figure*}
	% \subsection{Class Typical Signals}
	
\end{document}